\documentclass[prl,twocolumn,floatfix,showpacs]{revtex4}

\usepackage{psfig}

\begin{document}

\title{Observation of an unconventional
metal-insulator transition in overdoped CuO$_{2}$ compounds}

\author{F. Venturini$^1$, M. Opel$^1$, T. P. Devereaux$^2$, J. K.
Freericks$^3$, I.~T\"utt\H{o}$^4$, B. Revaz$^{5}$, E. Walker$^5$,
H. Berger$^6$, L. Forr\'o$^6$, and R. Hackl$^{1}$}

\affiliation{$^1$Walther Meissner Institute, Bavarian Academy of
Sciences, 85748 Garching, Germany} \affiliation{$^2$Department of
Physics, University of Waterloo, Waterloo, Ontario Canada N2L 3G1}
\affiliation{$^3$Department of Physics, Georgetown University,
Washington, DC 20057, U.S.A.} \affiliation{$^4$RISPO, Hungarian
Academy of Sciences, P.O.Box 49, 1525 Budapest, Hungary}
\affiliation{$^{5}$DPMC, University of Geneva, 1121 Gen\`eve,
Switzerland} \affiliation{$^6$EPFL, Ecublens, 1025 Lausanne,
Switzerland}

\date{\today}

\begin{abstract}
The electron dynamics in the normal state of
Bi$_2$Sr$_2$CaCu$_2$O$_{8+\delta}$ is studied by inelastic light
scattering over a wide range of doping. A strong anisotropy of the
electron relaxation is found which cannot be explained by
single-particle properties alone. The results strongly indicate
the presence of an unconventional quantum-critical metal-insulator
transition where ``hot'' (antinodal) quasiparticles become insulating while
``cold'' (nodal) quasiparticles remain metallic. A phenomenology is
developed which allows a quantitative understanding of the Raman
results and provides a scenario which links single- and
many-particle properties.
\end{abstract}


\maketitle

The normal state of copper-oxygen compounds is characterized by
several crossover lines separating regions of the phase diagram
with different physical properties. \cite{gutmann} The line
usually identified with the opening of a pseudogap at $T^{\ast}$
has been observed in many experiments which probe both
single-particle properties such as specific heat, angle-resolved
photoemission (ARPES), and many-particle properties such as NMR
and transport. \cite{tallon00,pseudo} An additional crossover at
higher temperatures and doping has been determined from anomalies
in NMR at  $T_0$. \cite{NMR} Moreover, another temperature scale
$T_{MI}$ associated with a metal-insulator transition has been
found in transport. \cite{highfield} Although several scenarios
have been proposed
\cite{zaanen89,SP,emery,zhang97,castellani97,varma,chakravarty}
there is no consensus of whether and how $T^{\ast},T_0$ and
$T_{MI}$ are related. It would be extremely useful if a
description of electron dynamics could be able to connect the
results from various experiments to a common origin.

From the point of view of critical phenomena it is not uncommon
that single-particle properties may show substantially different
behavior from many-particle properties. \cite{QPT} In the vicinity
of a quantum phase transition (QPT) single-particle properties,
e.g., density of states at the Fermi level may be uncritical while
two-particle properties such as the conductivity may deviate from
Fermi liquid behavior. For studying a putative QPT in planar
anisotropic systems like the copper-oxygen compounds a method is
desirable which not only probes many-particle properties but also
has resolution in {\bf k}~space. Raman scattering of light by
electrons can indeed meet these requirements in that the dynamical
response can be measured for different regions in the Brillouin
zone due to the possibility to independently adjust the
polarizations of incoming and outgoing photons. Thus, the $B_{1g}$
and $B_{2g}$ electronic Raman spectra can probe either ``hot''
(anti-nodal) or ``cold'' (nodal) electrons with momenta along the
principal axes and the diagonals of the CuO$_2$ plane,
respectively.

In this paper we present results from Raman scattering experiments
in Bi$_2$Sr$_2$CaCu$_2$O$_{8+\delta}$ (Bi2212) over a wide range
of effective doping $0.09 < p < 0.24$. We show that the transport
of ``hot'' electrons vanishes already at very high doping levels
while the ``cold'' ones still display metallic behavior. We
propose a phenomenology which can describe the results
quantitatively and outline how the various regimes in the phase
diagram could be connected.

The experiments have been performed on a set of single crystals
with different doping levels. We started from an extremely
homogeneous overdoped Bi2212 crystal grown by the traveling
solvent floating zone (TSFZ) method with a $T_c$ of 78 K and a
transition width of only 0.2 K. From this specimen several pieces
were cut and annealed at 620 K in oxygen partial pressures of 300
and 1350~bar to obtain transition temperatures of 62 and 56~K,
respectively.

Raw data of the Raman response
$\chi_{\mu}^{\prime\prime}(\omega,T,p)$ (with the symmetry index
$\mu=B_{1g},B_{2g}$) for various temperatures are plotted in Fig.
\ref{one}. Superimposed on the broad electronic continua are
narrow bands originating from Raman-active lattice vibrations. The
overall continua for $B_{2g}$ symmetry (bottom row) are relatively
doping independent and show a low-energy response, $\omega <
200~{\rm cm^{-1}}$, which decreases with increasing $T$. In
contrast, for $B_{1g}$ symmetry (top row) the continua show
non-trivial dependence on doping and $T$: the low-energy response
decreases with increasing $T$ in a similar way as in $B_{2g}$
symmetry for the strongly overdoped sample (Fig.~{\ref{one} (e)),
becomes temperature independent near optimum doping $p \geq 0.16$
(Fig.~{\ref{one} (c)), and increases with increasing $T$ for
samples below optimum doping (Fig.~{\ref{one} (a)). The different
doping dependences for the two symmetries become strikingly
evident if the spectra are plotted at a fixed temperature $T
\simeq 180~{\rm K}$ as a function of carrier concentration $p$ as
shown in Fig. \ref{longrange}. For $B_{2g}$ symmetry (Fig.
\ref{longrange} (b)) the doping dependence is very weak. For
$B_{1g}$ symmetry (Fig. \ref{longrange} (a)) the response is
suppressed strongly with decreasing doping in an energy range of
about 2000$~{\rm cm^{-1}}$ indicating the existence of a gap of
this magnitude for the ``hot'' electrons.

In order to link the results to momentum-dependent
electron dynamics we analyze the electronic Raman response in the
dc limit, $\omega \rightarrow 0$. For non-resonant scattering we
obtain \cite{tpd}
\begin{eqnarray}
\chi_{\mu}^{\prime\prime}(&\omega& \rightarrow  0 ) = \omega N_F
\times {} \nonumber\\
& &\times~\left<\gamma_{\mu}^{2}({\bf k}) \int d\xi
\left(-{\partial f^{0}\over{\partial \xi}}\right) {Z_{\bf
k}^{2}(\xi,T)\over{2\Sigma^{\prime \prime}_{\bf k}(\xi,T)}}
\right>. \label{chi0}
\end{eqnarray}
Here $N_F$ is the density of electronic levels at the Fermi energy
$E_F$ and $\gamma_{\mu}({\bf k})$ is the Raman scattering
amplitude, dependent upon incident (scattered) photon
polarizations $\hat e_{I(S)}$ corresponding to different
symmetries $\mu=B_{1g},B_{2g}$. $\Sigma^{\prime \prime}_{\bf k}$
is the imaginary part of the single-particle self energy related
to the electron lifetime as $\hbar/2\Sigma^{\prime \prime}_{\bf k}
(\omega,T) = \tau_{\bf k}(\omega,T)$, $Z_{\bf k}(\omega,T)$ is the
quasiparticle residue, $f^{0}$ is the Fermi distribution function,
and $\langle \cdots \rangle$ denotes an average over the Fermi
surface.

Using Eq.~(\ref{chi0}) we can define the Raman relaxation rate
via the slope of the spectra in the dc limit,
$\Gamma_{\mu}(T) \equiv N_F \langle \gamma_{\mu}^{2}({\bf k})
\rangle [\partial \chi_{\mu}^{\prime\prime}(\omega\rightarrow
0,T)/\partial \omega]^{-1}$. For an isotropic conventional metal
one obtains $\Gamma_{\mu}(T) = \hbar/\tau(T)$. In a correlated or
a disordered metal, however, a finite energy might be necessary to
move an electron from one site to another one. Thus, in spite of a
non-vanishing density of states at the Fermi level as observed in
an ARPES experiment for instance no current can be transported and
$\Gamma_{\mu}(T) \gg \hbar/\tau(T)$.

$\Gamma_{\mu}(T)$ can be explicitly extracted from the Raman
spectra by employing a memory or relaxation function approach.
\cite{opel2000,JCI} In Fig. \ref{two} (a) Raman relaxation rates
at a fixed temperature $T=200~{\rm K}$ obtained in this and
previous studies \cite{opel2000} are compiled. The magnitude of
$\Gamma_{B_{1g}}$ decreases by approximately 70\% for $0.09 \leq p
\leq 0.22$, while $\Gamma_{B_{2g}}$ is almost constant up to $p
\simeq 0.20$ and changes  by only 30\% in the narrow range $0.20 <
p < 0.22$. The abrupt crossover for $0.20 < p <0.22$ is
remarkable: the Raman relaxation rates rapidly decrease, and the
anisotropy vanishes. We emphasize that the changes for $0.20 < p
<0.22$ are observed for a set of samples prepared from a single
homogeneous piece and that the results agree well with those from
earlier experiments, hence the observed features are robust. The
variations of the relaxation rates with temperature $\partial
\Gamma_{\mu}(T)/ \partial T$ are shown in Fig. \ref{two}(b).
$\partial \Gamma_{B_{2g}}(T)/ \partial T$ (nodal electrons)
deviates only little from 2 in the entire doping range. The
logarithmic derivative $\partial[{\rm ln}\Gamma_{B_{2g}}(T)] /
\partial({\rm ln}T)$ [Fig. \ref{two}(c)] demonstrates that
$\Gamma_{B_{2g}}(T)$ varies essentially linearly with
temperature. The two properties combined yield
$\Gamma_{B_{2g}}(T) \simeq 2k_{B}T$. In contrast, both $\partial
\Gamma_{B_{1g}}(T)/
\partial T$ and $\partial ({\rm ln}\Gamma_{B_{1g}}(T))/
\partial({\rm ln}T)$ (anti-nodal electrons) are strongly
temperature dependent, increase continuously with $p$ and change
sign close to optimum doping. For $p \geq 0.22$ any kind of
anisotropy disappears.

It is both the apparent symmetry dependence of the Raman
relaxation rates, $\Gamma_{B_{2g}} < \Gamma_{B_{1g}}$, [Fig.
\ref{two} (a)] and their characteristic increase towards lower
temperature, $\partial \Gamma_{B_{1g}}(T)/
\partial T < 0$ for $p \leq 0.16$ [Fig. \ref{two} (b)] which
indicate that there is not only gap-like behavior but also a
strong anisotropy of the gap with the maxima located around the
M-points (see Fig. \ref{one} (e,f) and \ref{two} (a)). Thus the
``hot'' electrons show a crossover from metallic to insulating
behavior near optimum doping while the ``cold'' ones are metallic
for all doping levels at the temperatures examined. This is what
we call an anisotropic or generalized metal-insulator transition (MIT)
as opposed to a conventional Mott transition \cite{mott87} since
the charge excitations become gapped only on parts of the Fermi
surface, and the overall dc transport is still metallic.

The momentum dependence of the gap is reminiscent of both the
superconducting gap and the pseudogap \cite{tallon00,pseudo} being
compatible with $\mid d_{x^{2}-y^{2}} \mid$ symmetry. In spite of
a similar ${\bf k}$~dependence, however, an incipient
superconducting instability (pre-formed pairs) can be safely
excluded because of the high temperature (200~K) and doping level
($p>0.19$) of our experiment. The same holds for the pseudogap.
Its onset temperature $T^{\ast}$ actually merges with $T_c$
already for $0.16<p<0.19$. \cite{tallon00,pseudo} The scenario we
have in mind here is that of an anisotropic charge gap which
develops near the ``hot'' spots to minimize strong interactions
between electrons. \cite{patch} Therefore, we examine the effect
of an anisotropic normal-state gap on the Raman response.

An exact treatment of non-resonant electronic Raman scattering in
systems displaying a quantum-critical MIT in the limit of infinite
dimensions has been formulated for Hamiltonians displaying both
Fermi-liquid and non-Fermi-liquid ground states. \cite{JKFTPD}
However, the nature of a MIT in physical dimensions, the
development of an anisotropic gap, and their effect on the Raman
response is still an open issue. Rather than speculating on what
drives the QPT we consider here a phenomenological treatment for a
system near a QPT possessing a doping-dependent anisotropic gap in
the charge channel  $\Delta_{\rm C}(p)$. The result can be
directly derived from Eq. (\ref{chi0}) using the following
approximations: (1) we assume that the wavefunction
renormalization factor is constant, $Z_{\bf k}(\omega,T) \equiv
Z$, and non-zero only for $\Delta_{\rm C} \le~\mid 2\hbar\omega
\mid~\le E_{\rm b}$, e.g., for frequencies located in either part
of the band with a total width $E_{\rm b}$ separated symmetrically
with respect to the chemical potential by $\pm \Delta_{\rm C}/2$,
\cite{int} (2) we take the imaginary part of the self energy to be
momentum, energy, and doping independent, $\Sigma^{\prime
\prime}_{\bf k}(\omega,T) =\Sigma^{\prime \prime}(T)$, (3) we
assume a simple momentum dependence of the gap which is compatible
with the observed ${\bf k}$-dependence, $\Delta_{\rm C}({\bf k},p)
= \Delta_{\rm C}(\varphi,p) \equiv \Delta_{\rm
C}^{0}(p)\cos^2(2\varphi)$ with $\varphi$ the azimuthal angle on a
cylindrical Fermi surface, and (4) the doping dependence of the
gap is postulated to be proportional to $(1-p/p_c)^{\zeta}$ for $p
\leq p_c$. We obtain for $\Delta_{\rm C}(p), k_BT \ll E_{\rm b}$
\begin{equation}
{\Gamma_{\mu}(p,T)\over{2\Sigma^{\prime \prime}(T)}} =
\cases{\frac{1}{2}[1+\exp(\frac{\Delta_{\rm C}^ {\mu}(p)}{2k_BT})]
~{\rm (insulator),}\cr ~~~~~~~~1 ~~~~~~~~~~~~~~~{\rm (metal)}.}
\label{mott}
\end{equation}
$\Delta_{\rm C}^{\mu}(p)$ is a symmetry-specific effective gap
resulting from the Fermi-surface integration using
$\gamma_{B_{1g}}(\varphi) \sim \cos(2\varphi)$, and
$\gamma_{B_{2g}}(\varphi) \sim \sin(2\varphi)$.

In Fig.~\ref{two} along with the data, we plot theoretical curves
calculated from Eq.~({\ref{mott}}) using $\Sigma^{\prime
\prime}(T) = k_{B}T$ (being qualitatively compatible with ARPES
results in a wide range of doping \cite{arpesgamma}), $\Delta_{\rm
C}^{0}/k_{B} = 1100~{\rm K}$, and $\zeta = 0.25$.  For the
$B_{2g}$ symmetry which, according to our previous studies,
reflects dc and optical transport properties \cite{opel2000}, the
influence of the gap is weak, and the temperature dependence comes
essentially from $\Sigma^{\prime \prime}(T)$. The general trend of
the $B_{1g}$ rates, in particular the sign change, is well
reproduced by the phenomenology. It cannot be derived from
single-particle properties which, to our present knowledge, depend
only negligibly on doping in the range studied. \cite{kordyuk2001}
Due to the selection rules or, equivalently, the weighted Fermi
surface average (Eq.~({\ref{chi0}})) the MIT is particularly well
resolved in the Raman experiment and can be observed clearly up to
its onset at $p \simeq 0.22$.

This strongly suggests that an underlying quantum critical point
for this material lies at a doping of $p_{c} \simeq 0.22$ which is
considerably higher than $p_{c}^{tr} \simeq 0.19$ as derived from
transport properties. \cite{tallon00,pseudo,highfield}
Nevertheless, it appears that the two phenomena are directly
linked. The differences in the critical doping in transport and
Raman can readily be traced back to the selection rules and
directly reflect the unconventional anisotropic nature of the
transition. Since $p_{c} \simeq 0.22$ is also inferred from the
$T_0$~line \cite{gutmann} Raman scattering
probably captures the first onset
of non Fermi liquid behavior in this compound and can trace it
back to a correlation-induced localization of carriers. On the
other hand, the pseudogap at $T^{\ast}$ does not fit
straightforwardly into this scenario. It is either marks a pairing
or charge-ordering instability or is connected to $T_0$ in a more
complicated way through the electron-phonon
interaction. \cite{castellani97}

The existence of a QPT seems to be a general feature of the
cuprates although the critical doping depends on the material
class as demonstrated for low $T_c$ compounds. \cite{highfield}
Here we have shown that the QPT can be described
phenomenologically in terms of a generalized MIT with a strongly
anisotropic gap. In this framework the Raman data can be explained
quantitatively and reconciled with the single-particle results
from ARPES. Why $p_c$ is smaller
in systems with lower $T_c$'s remains an important question.\\

Many clarifying discussions with B.S. Chandrasekhar, D. Einzel and
A. Virosztek are gratefully acknowledged. F.V. is indebted to the
Gottlieb Daimler-Karl Benz Foundation for financial support. The
work is part of the DFG project under grant number HA2071/2-1.
J.K.F. acknowledges support from N.S.F. under grant number
DMR-9973225.

\newpage

\begin{figure*}
\caption[]{Raman response $\chi_{\mu}^{\prime \prime}(\omega,T,p)$
($\mu=B_{1g},B_{2g}$) of Bi$_2$Sr$_2$CaCu$_2$O$_{8+\delta}$. The
effective doping levels $p$ are derived from the empirical
relation $p = 0.16 \mp 0.11\sqrt{1-T_c/T_c^{\rm max}}$
{\protect\cite{tallon00}}. The spectra in the top row are measured
for $B_{1g}$ symmetry [polarizations of the incoming and outgoing
light perpendicular and at $45^{\circ}$ to the copper- (full)
oxygen (open symbols) bonds]. For this configuration, the
sensitivity is highest around the M points and vanishes along the
diagonals of the Brillouin zone. The Fermi surface is indicated as
bold line [see inset of (e)]. For $B_{2g}$ symmetry (bottom row),
the centers of the quadrants are projected out while the principal
axes become invisible [inset of (f)].} \label{one}
\end{figure*}

\begin{figure*}
\caption[]{Raman spectra $\chi_{\mu}^{\prime \prime}(\omega,p)$ as
a function of doping on an extended energy scale. For clarity the
contributions from lattice vibrations have been subtracted out. }
\label{longrange}
\end{figure*}

\begin{figure*}
\caption[]{(a) Raman relaxation rates $\Gamma_{\mu}(p)$ as a
function of doping $p$. The smooth lines are fits to the data
employing Eq.~({\ref{mott}}) with a {\bf k}-dependent gap with the
maxima close to the M points (inset). The metallic part above
$p_c=0.22$ is shaded. The vertical line corresponds to optimum
doping. (b) Variation with temperature of the relaxation rate,
$\partial\Gamma_{\mu}(T)/
\partial T$. The smooth lines are
theoretical predictions. (c) Logarithmic derivatives of the Raman
relaxation rates indicating power-law behavior in the temperature
dependence.} \label{two}
\end{figure*}


\addcontentsline{toc}{section}{Bibliography}
\begin{thebibliography}{99}

\bibitem{gutmann}
M. Gutmann, E.S. Bo\u{z}in, S.J.L. Billinge, cond-mat 0009141
(2000).
\bibitem{tallon00}
J.L. Tallon and J.W. Loram, Physica C {\bf 349}, 53 (2001).
\bibitem{pseudo}
T. Timusk and B.W. Statt, Rep. Prog. Phys. {\bf 62}, 61 (1999).
\bibitem{NMR}
H. Alloul, T. Ohno, and P. Mendels, Phys. Rev. Lett. {\bf 63},
1700 (1989).
\bibitem{highfield}
Y. Ando {\it et al.}, Phys. Rev. Lett. {\bf 75}, 4662 (1995); {\bf
77}, 2065 (1996); {\bf 79}, 2595 (1997); Phys. Rev. B {\bf 56},
R8530 (1997); G. S. Boebinger {\it et al.}, Phys. Rev. Lett. {\bf
77}, 5417 (1996); P. Fournier {\it et al.}, {\it ibid} {\bf 81},
4720 (1998); S. Ono {\it et al.}, {\it ibid} {\bf 85}, 638 (2000).

\bibitem{zaanen89}
J. Zaanen and O. Gunnarsson, Phys. Rev. B {\bf 40}, 7391 (1989).
\bibitem{SP}
A. Sokol and D. Pines, Phys. Rev. Lett. {\bf 71}, 2813 (1993).
\bibitem{emery}
V. Emery and S. Kivelson, Nature {\bf 374},434 (1995).
\bibitem{zhang97}
S.C. Zhang, Science {\bf 275}, 1089 (1997).
\bibitem{castellani97}
S. Andergassen {\it et al.}, Phys. Rev. Lett. {\bf 87}, 56401
(2001).
\bibitem{varma}
C.M. Varma, Phys. Rev. B {\bf 61}, R3804 (2000)
\bibitem{chakravarty}
S. Chakravarty {\it et al.}, Phys. Rev. B {\bf 63}, 094503 (2001).

\bibitem{QPT}
S. Sachdev, Science {\bf 288}, 475 (2000).
\bibitem{tpd}
T.P. Devereaux and A.P. Kampf , Phys. Rev. B {\bf 59}, 6411
(1999).
\bibitem{opel2000}
M. Opel {\it et al.}, Phys. Rev. B {\bf 61}, 9752
(2000).
\bibitem{JCI}
J.G. Naeini {\it et al.}, Can. J. Phys. {\bf 5-6}, 483
(2000).
\bibitem{mott87} N. Mott, {\it Conduction in
Non-Crystalline Materials} (Clarendon Press, Oxford, 1987).

\bibitem{patch}
N. Furukawa {\it et al.}, Phys. Rev. Lett. {\bf 81}, 3195 (1998);
J. Gonz\'alez {\it et al.}, Phys. Rev. Lett. {\bf 84}, 4930
(2000).
\bibitem{JKFTPD}
J. K. Freericks and T. P. Devereaux, Phys. Rev. B {\bf 64}, 125110
(2001); J. K. Freericks, T. P. Devereaux, and R. Bulla, Phys. Rev.
B {\bf 64}, 233114 (2001).
\bibitem{int}
As shown for a marginal Fermi liquid, for instance, $Z=0$ does not
necessarily imply that one cannot observe a Fermi surface. The
approximation is therefore not in conflict with ARPES. In general,
the gap $\Delta_{\rm C}$ should be considered an activation energy
for moving a particle rather than a single particle gap.

\bibitem{arpesgamma}
M.R. Norman, M. Randeria, H. Ding, and J.C. Campuzano, Phys. Rev.
B {\bf 57}, R11093 (1998); T. Valla, A.V. Fedorov, P.D. Johnson,
Q. Li, G.D. Gu, and N. Koshizuka, Phys. Rev Lett. {\bf 85}, 828
(2000).
\bibitem{kordyuk2001}
A.A. Kordyuk {\it et al.}, cond-mat/0201485 (2001).


\end{thebibliography}
\end{document}